# Magnetic and Electrical Properties of Ordered 112-type Perovskite LnBaCoMnO$_{5+\delta}$ (Ln = Nd, Eu)

Asish K. Kundu[1,2*], V. Pralong[2], B. Raveau[2] and V. Caignaert[2]

[1]*Indian Institute of Information Technology Design & Manufacturing, Dumna Airport Road, Jabalpur –482005, India*

[2]*Laboratoire CRISMAT, ENSICAEN UMR6508, 6 Boulevard Maréchal Juin, Cedex 4, Caen-14050, France*

## ABSTRACT

Investigation of the oxygen-deficient 112-type ordered oxides of the type LnBaCoMnO$_{5+\delta}$ (Ln = Nd, Eu) evidences certain unusual magnetic behavior at low temperatures, compared to the LnBaCo$_2$O$_{5+\delta}$ cobaltites. One observes that the substitution of manganese for cobalt suppresses the ferromagnetic state and induces strong antiferromagnetic interactions. Importantly, NdBaCoMnO$_{5.9}$ depicts a clear paramagnetic to antiferromagnetic type transition around 220 K, whereas for EuBaCoMnO$_{5.7}$ one observes an unusual magnetic behavior below 177 K which consists of ferromagnetic regions embedded in an antiferromagnetic matrix. The existence of two sorts of crystallographic sites for Co/Mn and their mixed valence states favor the ferromagnetic interaction whereas antiferromagnetism originates from the Co$^{3+}$-O-Co$^{3+}$ and Mn$^{4+}$-O-Mn$^{4+}$ interactions. Unlike the parent compounds, the present Mn-substituted phases do not exhibit prominent magnetoresistance effects in the temperature range 75-400K.

(Some figures in this article are in colour only in electronic version)

---

*For correspondence: *e-mail:* asish.k@gmail.com/asish.kundu@iiitdmj.ac.in

# I. INTRODUCTION

Ordered 112-type perovskite cobaltites are increasingly recognized as materials of importance due to rich physics and chemistry in their layered structure [1, 2]. Apart from colossal magnetoresistance effect, like manganites, the different form of cobaltites exhibit interesting phenomena including magnetic ordering, electronic phase separation, insulator-metal transition, large thermoelectric power at low temperature [1]. A few of these behaviors are of great interest because of their potential applications as read heads in magnetic data storage, oxidation catalyst, gas sensors etc, and also in other applications depending upon their particular properties [1, 2]. The well known oxygen deficient cobalt perovskite $LnBaCo_2O_{5+\delta}$ (Ln = rare earth) is basically derived from the 112-type ordered $YBaFeCuO_{5+\delta}$ structure [3]. In the case of cobaltites for $\delta = 0$ the structure only consists of double pyramidal cobalt layers, whereas for $\delta > 0$ there will be $CoO_6$ octahedra as well as $CoO_5$ pyramidal layers containing the barium cations, interleaved with rare earth layers [1]. The discovery of large magnetoresistance (MR) in this type of structure has renewed great interest in 112-ordered cobaltites since 1997 [1]

Likewise, the existence of ordered oxygen deficient perovskites $LnBaCo_{2-x}M_xO_{5+\delta}$ (Ln = rare earth and M = metal cations) with interesting magnetic properties has stimulated the research of doped 112-ordered cobaltites [4]. In fact, the magnetic and electron transport properties of this type of oxides are very sensitive to their oxygen stoichiometry and to their complex crystal chemistry [5]. Numerous studies performed on the doped 112-ordered cobaltites have shown that their fascinating physical behavior is complex, in connection with their large possibility for cation ordered-disordered phenomena and oxygen deficiency [4, 5]. According to our knowledge for such systems, with mixed Co & Mn-cations at the B-site of the perovskite, only two compounds are reported till date i.e. YBaMnCoO$_5$ [6] and NdBaMnCoO$_{5+\delta}$ [7]. The latter one (with $\delta$ = 0, 1) has been investigated recently by Snedden et al [7], who reported a tetragonal structure with P4/mmm space group (unit cell $a_p$ x $a_p$ x $2a_p$) at room temperature. Moreover, the neutron diffraction studies show a G-type antiferromagnetic (AFM) structure for $\delta$ = 0 but there is no evidence for magnetic ordering at low temperature for $\delta$ = 1, although a weak ferromagnetic (FM) ordering around 228 K was predicted for the latter one. In fact, no detailed investigation has allowed the FM ordering to be confirmed, so that it may be due to the presence of impurity phases [7]. There is certainly lack of supportive information on the FM phase below 228 K. Bearing these results in mind, we have investigated the effect of Mn-doping at the cobalt site in 112-ordered cobaltite's. In contrast, the magnetic data reported for NdBaCoMnO$_6$ [7], we have defined the transition as AFM type and explored the low temperature region by means of isothermal magnetization to confirm the magnetic interaction. Herein, we also report a new 112-ordered phase EuBaCoMnO$_{5.7}$ that shows a magnetic transition at 177 K, with phase separation between an AFM matrix and FM domains below this temperature.

## II. EXPERIMENTAL PROCEDURE

The ordered LnBaCoMnO$_{5+\delta}$ (Ln = Nd, Eu) perovskites were synthesized by means of a soft-chemistry method. Stoichiometric amounts of metal oxides and nitrates Ln$_2$O$_3$, Ba(NO$_3$)$_2$-H$_2$O, Co(NO$_3$)$_2$-6H$_2$O and Mn(NO$_3$)$_2$-4H$_2$O were dissolved into distilled water and citric acid was added to the solution in the molar ratio. After adding citric acid into the solution, the mixture solutions were heated at 100 °C for few hours and evaporated at 150 °C to form an amorphous dry gel, which was decomposed at 800 °C for overnight to burn away the carbon and nitrogen residues. The powder samples were ground thoroughly and pressed into rectangular bars, and finally sintered at 1200-1320 °C in Ar flow (5 N) for 36 h in plate-type platinum crucible. Heating and cooling rates were kept slow (2 °C/min) to enable better *A*-cation and oxygen vacancy ordering. The parent compounds $LnBaCo_2O_{5+\delta}$ (Ln = Nd, Eu) were prepared by the conventional solid state reaction method as reported in the literature [1].

Small parts of the sintered bar were taken and ground to form fine powder to record the X-ray diffraction (XRD) pattern, using a Philips diffractometer employing Cu-K$\alpha$ radiation. The phases were identified by performing Rietveld [8] analysis in the 2$\theta$ range of 5°-120° and the lattice parameters were calculated accordingly, listed in Table I. Composition analysis was carried out by energy dispersive spectroscopy (EDS) analysis using a JEOL 200CX scanning electron microscope, equipped with a KEVEX analyzer. The oxygen stoichiometry was determined by iodometric titrations. The error in oxygen content was ± 0.05.

Other pieces of the rectangular bars were taken for magnetization, resistivity and thermopower measurements. A Quantum Design physical properties measurements system (PPMS) magnetometer was used to investigate the magnetic properties of the samples. The temperature dependence of the zero-field-cooled (ZFC) and field-cooled (FC) magnetization was measured in different applied magnetic fields. Hysteresis loops M(H) were recorded at different temperatures. In the measurements of the ZFC magnetization, the sample was cooled from 300 K to 10 K in zero-field, the field was applied at 10 K and the magnetization recorded on re-heating the sample. In the FC measurements the sample was cooled (from 300 K) in the applied field to 10 K and the magnetization recorded on re-heating the sample, keeping the field applied. The electrical measurements were carried out on a rectangular-shaped (6.90x2.65x2.20mm$^3$) sample by a standard four-probe method in the temperature range of 10-400 K. The electrodes on the sample were prepared by ultrasonic deposition method using indium metal.

## III. RESULTS AND DISCUSSION

### A. Structural analysis at room temperature

All the perovskite based 112-ordered samples LnBaCo$_{2-x}$Mn$_x$O$_{5+\delta}$ (with Ln = Nd/Eu and x = 0, 1), confirm the single-phase, without any traces of impurities as shown in FIG. 1. The crystal structure adopted by ordered cobaltite LnBaCo$_2$O$_{5+\delta}$ (Ln = rare earth) has been reported to be either tetragonal P4/mmm (a$_p$×a$_p$×2a$_p$), orthorhombic Pmmm (a$_p$×2a$_p$×2a$_p$ or a$_p$×a$_p$×2a$_p$), or orthorhombic Pmmb (a$_p$×2a$_p$×2a$_p$). Here a$_p$ refers to the basic cubic perovskite cell parameters (ca. 3.9Å). The doubling in c parameter is due to the ordering of Ln and Ba into layers. The cells doubling in b, and the transition from tetragonal to orthorhombic, have been suggested to arise from different orderings between oxygen and vacancies in the [LnO$_\delta$] layer [1]. The obtained crystal structure for the phases NdBaCo$_2$O$_{5.6}$, NdBaCoMnO$_{5.9}$, EuBaCo$_2$O$_{5.5}$ and EuBaCoMnO$_{5.7}$, along with the oxygen content are in good agreement to the literature data [1].

### B. Magnetic properties of the parent compounds LnBaCo$_2$O$_{5.5+\delta}$ revisited

FIG. 2 shows the ZFC and FC magnetization, M (T), for NdBaCo$_2$O$_{5.6}$ and EuBaCo$_2$O$_{5.5}$ in an applied field of 1000 Oe which confirms the successive transitions from a PM state to a FM state and then to AFM state (mainly for Eu-phase), as previously reported in the literature [1]. The existence of ferromagnetism for EuBa{Co$^{III}$}$_2$O$_{5.5}$ is in perfect agreement with the possibility of FM interactions due to intermediate spin state $Co^{3+}$ species (with the existence of two kinds of crystallographic sites) and for $NdBa\{Co^{III}\}_{1.8}\{Co^{IV}\}_{0.2}O_{5.6}$ also between $Co^{3+}$ and $Co^{4+}$ species, since the average oxidation state of cobalt is close to 3.10. Moreover, for $NdBaCo_2O_{5.6}$ in the low temperature regime several magnetic transitions noticed particularly in ZFC data a FM to AFM-type transition appeared around 210K, but in FC data that transition is not so prominent although a small kink is observed. In the case of $EuBaCo_2O_{5.5}$ the AFM transition ($T_N \sim 245K$) is clearly observed in both ZFC-FC data, yet the divergence between them persists down to low temperature. Hence, for both systems, the magnetization value is non-zero below the AFM transition and there is a large irreversibility between the ZFC-FC magnetization data even at higher fields. This signifies some kind of short range FM ordering in the low temperature AFM region. The isothermal, M(H), curves studied at different temperatures (insets of Fig. 2), also confirm this short-range FM ordering. One indeed observes a clear hysteresis loop for both compounds at 10 K and the hysteresis is rather smaller for $EuBaCo_2O_{5.5}$ compared to $NdBaCo_2O_{5.6}$, with remanent magnetization ($M_r$) values of 0.024 and 0.16 $\mu_B$/f.u. and coercive fields ($H_C$) of $\sim$ 4.1 and 9.5 kOe respectively. The highest value of magnetic moment is only $\sim$ 0.21 $\mu_B$/f.u. (0.87 $\mu_B$/f.u.) for Eu (Nd)-compound, which is less than the spin-only value of $Co^{3+}$-ions in the intermediate spin (IS) state. Below the magnetic ordering temperature (for both phases) the obtained hysteresis loop signifies a typical FM state and at higher temperatures (T > 200 K) the M(H) behavior is linear, corresponding to a PM state. Another interesting feature at low temperature is the unsaturated behavior of the M(H) curve even at higher fields, which is in agreement with electronic phase separation as pointed out by several authors [1] and will not be discussed here.

## C. Magnetic properties of the Mn-doped phases: LnBaCoMnO$_{5+\delta}$

The LnBaCoMnO$_{5+\delta}$ (Ln = Nd, Eu) phases show a significant different transition temperature compared to their parent oxides LnBaCo$_2$O$_{5+\delta}$ [1] and LnBaMn$_2$O$_{5+\delta}$ [9].

*Magnetic behavior of NdBaCoMnO$_{5.9}$*

FIG. 3 exhibits the temperature dependent ZFC and FC magnetization curves recorded at 1000 Oe for NdBaCoMnO$_{5.9}$. These are almost similar to that of NdBaCoMnO$_6$ reported by Snedden et al [7]. In contrast to the report of the authors, in the temperature range of 10-400 K the system exhibits a clear PM to AFM transition around T$_N$ ~ 220K (shown by arrow mark in the figure). However, the magnetization value does not become zero below T$_N$, as expected for an AFM system and the value remains almost constant in the low temperature AFM state for both ZFC-FC data. Moreover, there is no such magnetic irreversibility between ZFC and FC data below T$_N$ unlike the parent phase NdBaCo$_2$O$_{5.6}$. However, the magnetization value increases rapidly at low temperature (T < 80 K), which may be due to the PM contribution from the Nd-cation as reported in the literature for some of the lanthanides [1, 9]. Inset of FIG. 3 exhibits the inverse magnetic susceptibility versus temperature plot throughout the measured range. The latter follows a simple Curie-Weiss law in the 220 ≤ T ≤ 400 K range and yields a PM Weiss temperature (θ$_p$) of 11 K and an effective magnetic moment (μ$_{eff}$) of ~ 7.59 μ$_B$/f.u. The obtained value of θ$_p$ is positive and much lower than the T$_N$ value (220 K), suggesting FM interactions in the high temperature region. This indicates that due to Mn-substitution at the Co-site the FM interactions disappear completely at low temperature, yet weak contributions are still present at high temperature. We have further studied the isothermal magnetization behavior, M(H), of the ZFC sample recorded below the AFM transition to establish this behavior. FIG. 4 shows the M(H) curves at 10 and 100 K, exhibiting a typical feature of an AFM state below $T_N$ and above this temperature the M(H) behavior is also linear corresponding to a PM state. Considering the potential of the couples $Co^{4+}/Co^{3+}$ and $Co^{3+}/Co^{2+}$ with respect to the couple $Mn^{4+}/Mn^{3+}$, it clearly appears that $Co^{3+}$ is reduced into $Co^{2+}$ in the presence of $Mn^{3+}$ according to the equation $Co^{3+} + Mn^{3+} \rightarrow Co^{2+} + Mn^{4+}$ [10]. As a result, the charge balance in this oxide can be formulated as $NdBa\{Co^{III}\}_{0.8}\{Co^{II}\}_{0.2}Mn^{IV}O_{5.9}$. Thus the majority of interactions $Mn^{4+}$-O-$Mn^{4+}$, $Co^{3+}$-O-$Co^{3+}$, and $Co^{3+}$-O-$Mn^{4+}$ are predicted to be antiferromagnetic at low temperature, according to Goodenough-Kanamori rules [10]. $Co^{2+}$-O-$Mn^{4+}$ interaction might be ferromagnetic, but this requires an ordering of the $Co^{2+}$ and $Mn^{4+}$ species, as reported for the double perovskite $La_2MnCoO_6$ [11], which is not the case for the present compound. The $Co^{2+}$-O-$Co^{3+}$ interactions may be either ferro- or antiferromagnetic, but in any case, they should bring a very low contribution, due to the low $Co^{2+}$ content. This behavior is different from parent $NdBaCo_2O_{5.6}$ compound, which exhibits FM interaction between $Co^{3+}$ and $Co^{4+}$ ions according to Goodenough-Kanamori rules [10].

*Magnetic behavior of EuBaCoMnO$_{5.7}$*

Bearing in mind the complex magnetic properties observed for $EuBaCo_2O_{5.5}$ and the strong AFM interactions in $NdBaCoMnO_{5.9}$, the possibility of a similar magnetic interaction with mixed Co-Mn system $EuBaCoMnO_{5.7}$ has been investigated in details. For as prepared (in argon atmosphere) sample a large oxygen deficiency is observed, leading to the chemical formula $EuBa\{Co^{III}\}_{0.4}\{Co^{II}\}_{0.6}Mn^{IV}O_{5.7}$. The temperature dependent ZFC-FC magnetization, M(T), is measured in the range of 10-400 K under the applied fields of 100, 1000 & 5000 Oe (FIG.5). With decreasing temperature the FC magnetization curve depicts a steep increase of the magnetization around 177 K followed by a small kink at 130 K. The FC data suggest a PM to FM-like transition $T_C \sim$ 177 K, yet the highest obtained moment is only 0.06 $\mu_B$/f.u. (at H = 5000 Oe) and there is no saturation in the magnetization data down to low temperature. Moreover the ZFC-FC curves exhibit a significant thermo-magnetic irreversibility below $T_C$ even at higher fields. It must also be emphasized that though the system becomes FM-like below 177 K, the magnitude of the magnetic moment as well as the ZFC-FC behavior (at higher fields) do not justify the long-range FM ordering or a true FM behavior. Indeed, the parent manganite $EuBaMn_2O_5$ phase also exhibits a magnetic transition around 150 K, yet for $EuBaMn_2O_6$ the transition is around 260 K. The magnetic state for $EuBaMn_2O_6$ is explained by inhomogenous ferromagnetism, whereas for $EuBaMn_2O_5$ as FM type [9]. In order to clarify the FM-like state in $EuBaCoMnO_{5.7}$ the isothermal magnetization behavior, M(H), at three different temperatures has been investigated as shown in FIG. 6. The M(H) curve at 10 K depicts a prominent hysteresis loop with a remanent magnetization, $M_r$, and coercive field, $H_C$, values of 0.03 $\mu_B$/f.u. and 2.4 kOe respectively indicating a FM-like state. The hysteresis loop persists up to 100 K (see FIG. 6), though the $H_C$ (~ 0.4 kOe at 100 K) decreases with increasing temperature and above $T_C$ the M(H) behavior is linear akin to PM state (T= 300K). Nevertheless, the maximum value of the magnetic moment measured in 50 kOe is only 0.5 $\mu_B$/f.u. at 10 K, which is much smaller than the value expected for Co and Mn ions. Therefore the weak FM-like feature of the present compound may be due to the canting of the magnetic spin alignment in the G-type AFM structure at low temperature as pointed out by several authors for 112-ordered cobaltites [1]. It is worth pointing out that at low temperature the magnetization value does not saturate even at higher fields. This is due to a superposition of a quite large AFM state and a small region of FM state similar to the phases $EuBaCo_{1.92}M_{0.08}O_{5.5-\delta}$ with M = Zn, Cu [4]. Hence, the obtained smaller value of magnetic moment for $EuBaCoMnO_{5.7}$ below $T_C$ can be explained by the appearance of electronic phase separation i.e. the presence of small FM domains inside the AFM matrix. Indeed, in this phase the $Co^{2+}$ content has increased with respect to the $NdBaCoMnO_{5.9}$ phase, and the number of $Co^{2+}$-O-$Co^{3+}$ interactions is significantly larger. Thus, if the latter are FM at low temperature, as observed for $EuBaCo_2O_{5.33}$ [Ref.1; Seikh et al] this could explain that there is a strong competition between positive $Co^{2+}$-O-$Co^{3+}$ FM interactions and negative $Mn^{4+}$-O-$Mn^{4+}$ and $Co^{2+}$-O-$Co^{2+}$ AFM interactions, whereas the FM interactions would dominate over AFM at higher fields. But, the contribution of AFM interaction is not negligible; hence the isothermal M(H) exhibits an unsaturated hysteresis behavior supporting the electronic phase separation model [12].

**D. Electron transport properties**

The temperature dependence of the electrical resistivity (ρ) behavior for $LnBaCo_{2-x}Mn_xO_{5+\delta}$ (Ln = Nd, Eu and x = 0, 1) samples is shown in FIG. 7. The parent cobaltites show the well known insulator-metal transition in the proximity of 350 K as reported in the literature [1]. In contrast for the Mn-substituted phases the insulator-metal transition completely disappears throughout the temperature range 60-400 K quite similar to $YBaMnCoO_5$ phase [6]. With decreasing temperature the resistivity increases for $NdBaCoMnO_{5.9}$ and $EuBaCoMnO_{5.7}$ samples and the value is very high at low temperature, crossing the instrument limitations below 60 K. The rapid increase in temperature coefficient of resistivity (dρ/dT) from room temperature to low temperature signifies the insulating behavior. Thus, none of Mn-doped samples show an insulator-metal transition like the parent phases. The highest value of the resistivity at room temperature is observed for EuBaCoMnO$_{5.7}$. This is due to the smaller A-site cation radius for EuBaCoMnO$_{5.7}$ compared to NdBaCoMnO$_{5.9}$. With decreasing the rare earth ion size the band gap between the valence and conduction band increases, as a result resistivity increases. At this point, it is important to mention that since the EuBaCoMnO$_{5.7}$ shows a prominent magnetic ordering below 177 K, so we have also investigated the magnetoresistance effect in an applied field of 70 kOe. In a magnetic field, ρ(T) does not show a large change in the resistivity behavior, in contrast to parent cobaltites, except a slight decrease in the magnitude is noticed at low temperature. Therefore we have calculated the magnetoresistance (MR) as, MR (%) = [{ρ(7)-ρ(0)}/ρ(0)]x100, where ρ(0) is the sample resistivity at 0 T and ρ(7) in an applied field of 70 kOe. The parent EuBaCo$_2$O$_{5.5}$ shows MR in the range 10-20 % in the measured temperature range as expected [1], but the NdBaCoMnO$_{5.9}$ and EuBaCoMnO$_{5.7}$ compounds exhibit rather small MR value. In fact for NdBaCoMnO$_{5.9}$ the MR value is negligibly small (around -2 % ) even at low temperature, whereas EuBaCoMnO$_{5.7}$ exhibits a MR value close to -5 % (at 80K). The evidence of small negative MR for EuBaCoMnO$_{5.7}$ at low temperatures is considered to be related to the suppression of spin dependent scattering of the electrons below the magnetic ordering temperature with the application of magnetic field.

## IV. CONCLUSIONS

Oxygen deficient 112-type ordered NdBaCoMnO$_{5.9}$ and EuBaCoMnO$_{5.7}$ perovskites were synthesized by soft-chemistry method. At room temperature there is perfect ordering between A-site cations, whereas the B-site cations are distributed randomly, in agreement with previous investigations [4, 7]. The valence states of Co/Mn-cations at the B-site are most probably in Co$^{2+}$, Co$^{3+}$ and Mn$^{4+}$ mixed states and majority of them interact antiferromagnetically whereas the presence of FM interaction cannot be ruled out at low temperature. As a matter of fact the prominent FM T$_C$, insulator-metal transition and the large MR effect disappear completely for the Mn-substituted phases, in contrast to the parent cobaltites. However, the nature of magnetic interactions at low temperature varies with the nature of rare earth ions. NdBaCoMnO$_{5.9}$ is AFM below room temperature whereas EuBaCoMnO$_{5.7}$ sample shows weak ferromagnetism as well as MR effects at low temperature due to electronic phase separation. The system appears to be phase separated into FM clusters and AFM matrix. The M (H) measurements also support this assumption. In addition the electron transport properties of both Mn-doped compounds reveal the presence of insulating phase at low temperature irrespective of their magnetic behavior.

## ACKNOWLEDGEMENTS

The authors gratefully acknowledge the CNRS and the Ministry of Education and Research for financial support. AKK thanks Prof A. Ojha for faculty research grants.

**Table 1:** Crystallographic Data for (a) $NdBaCo_2O_{5.6}$ (b) $NdBaCoMnO_{5.9}$ and (c) $EuBaCoMnO_{5.7}$

| Compound | $NdBaCo_2O_{5.6}$ | $NdBaCoMnO_{5.9}$ | $EuBaCoMnO_{5.7}$ |
|---|---|---|---|
| Crystal system | Orthorhombic | tetragonal | orthorhombic |
| Space-group | P m m m (47) | P 4/m m m (123) | P m m m (47) |
| Cell parameters | a=3.900(1)Å b=7.826(2) Å c=7.611(1) Å | a=3.893(2) Å b=7.776 (2)Å c=7.697(1) Å | a=3.884(2)Å b=3.893(2) Å c=7.653(1) Å |
| Cell volume | 232.32 (2) Å$^3$ | 116.54(2) Å$^3$ | 115.70(3) Å$^3$ |
| Z | 2 | 1 | 2 |
| $\chi^2$ | 2.07 | 3.22 | 7.10 |
| $R_B$ | 9.71% | 6.42% | 8.93% |

**FIG. 1.** Rietveld analysis of XRD pattern for (a) NdBaCo$_2$O$_{5.6}$ (b) NdBaCoMnO$_{5.9}$ and (c) EuBaCoMnO$_{5.7}$ at room temperature. Open symbols are experimental data and the dotted, solid and vertical lines represent the calculated pattern, difference curve and matched profile respectively.

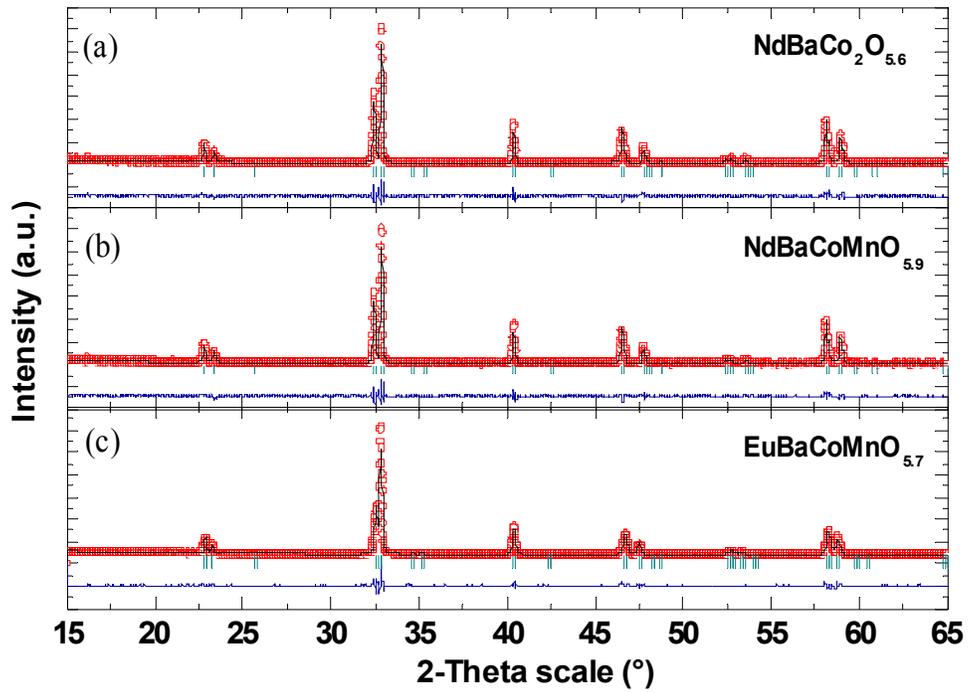

**FIG. 2.** Temperature dependent ZFC (open symbol) and FC (solid symbol) magnetization, M, of (a) $NdBaCo_2O_{5.6}$ and (b) $EuBaCo_2O_{5.5}$ (H = 1000 Oe). The insets show typical hysteresis curves at different temperatures.

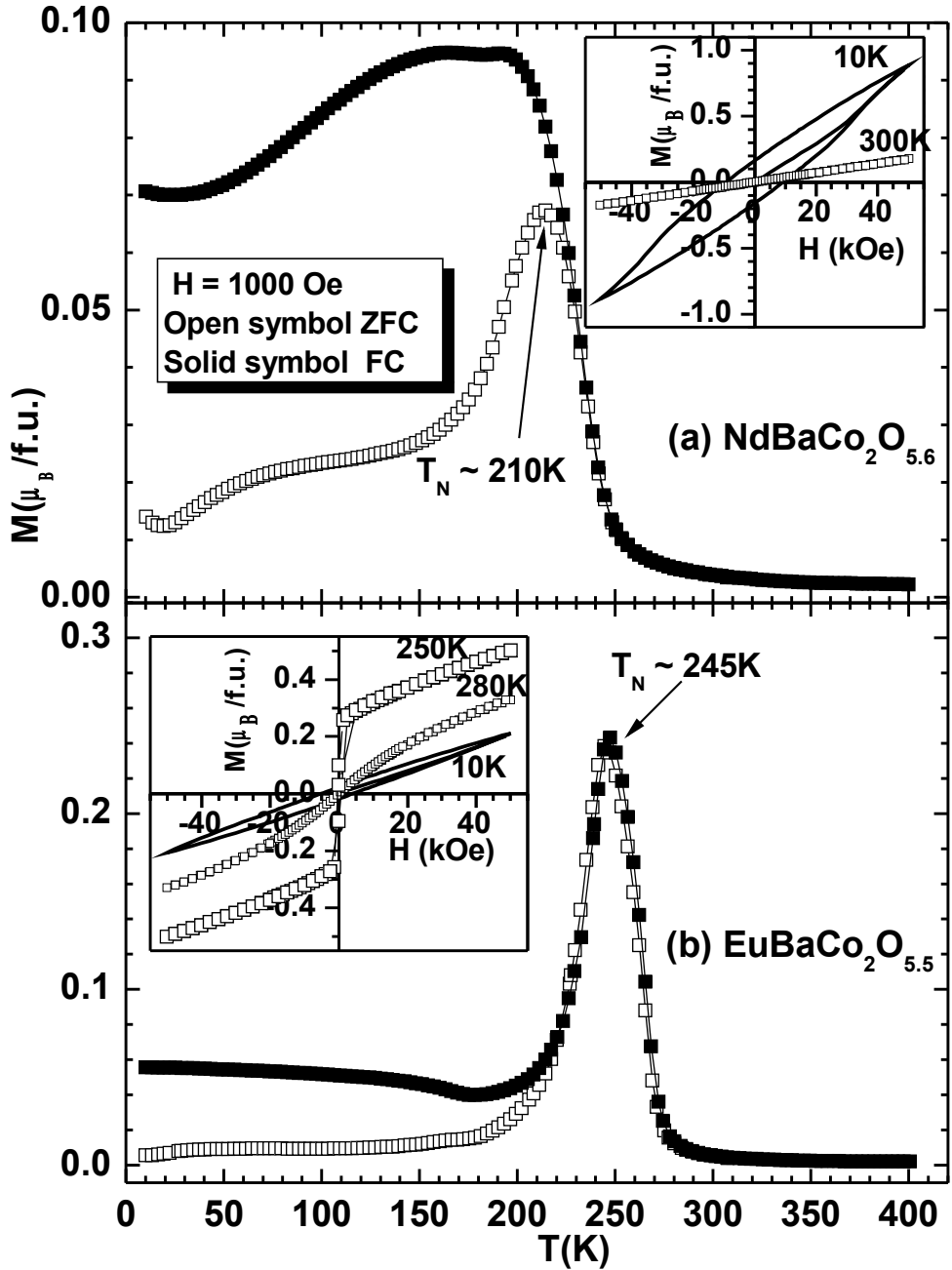

**FIG. 3.** Temperature dependent ZFC (open symbol) and FC (solid symbol) magnetization, M, of NdBaCoMnO$_{5.9}$ in an applied fields of 1000 Oe. Inset figure shows the inverse susceptibility, $\chi^{-1}$, vs temperature plot.

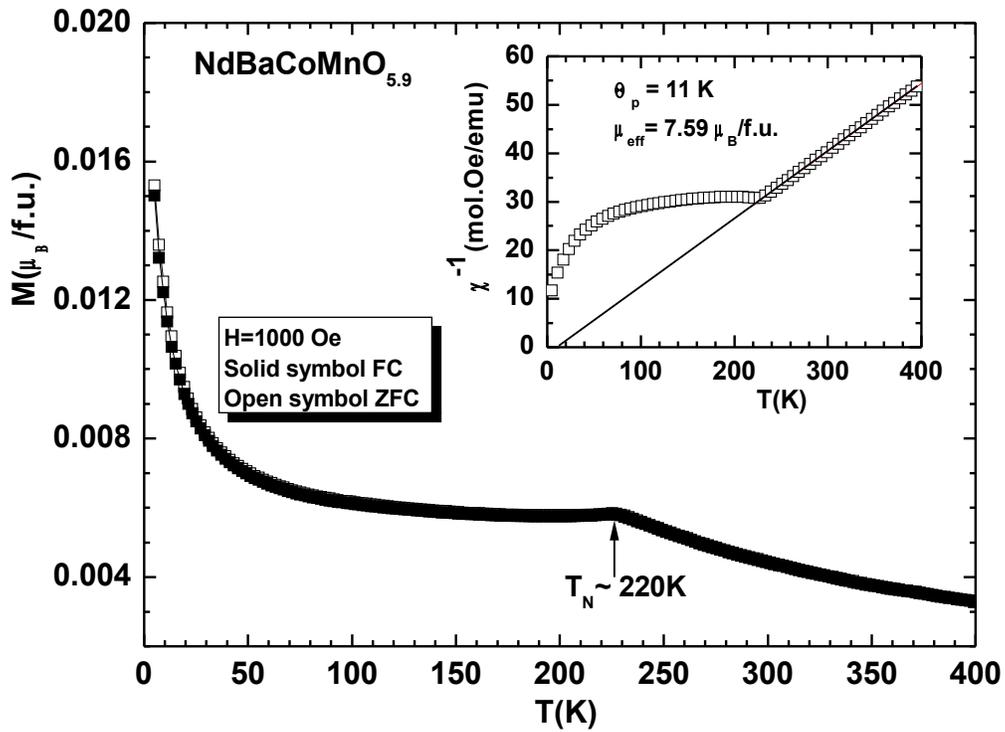

**FIG. 4.** Magnetic field dependence of isothermal magnetization, M(H), at two different temperatures for NdBaCoMnO$_{5.9}$.

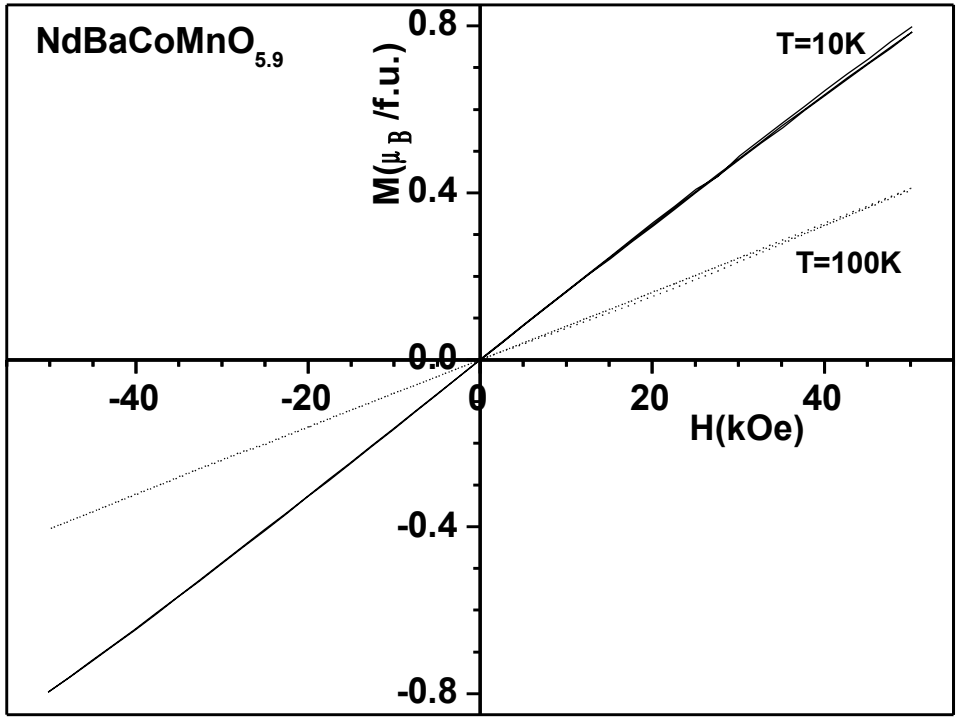

**FIG. 5.** Temperature dependent ZFC (open symbol) and FC (solid symbol) magnetization, M, of EuBaCoMnO$_{5.7}$ at different applied fields (a) H = 100 (b) H = 1000 (inset shows inverse susceptibility, $\chi^{-1}$, vs temperature plot) and (b) H = 5000 Oe.

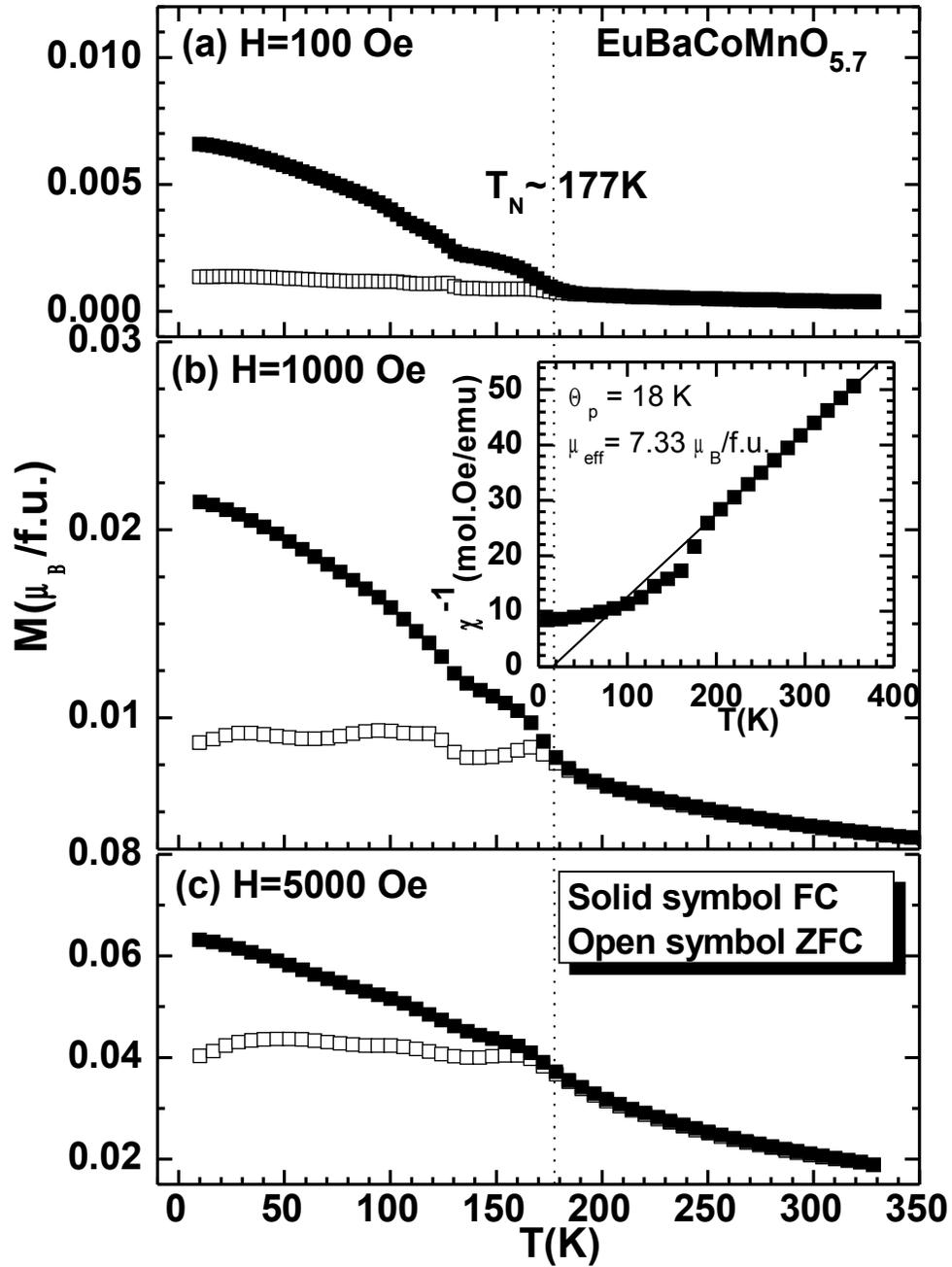

**FIG. 6.** Magnetic field dependence of isothermal magnetization at three different temperatures for EuBaCoMnO$_{5.7}$.

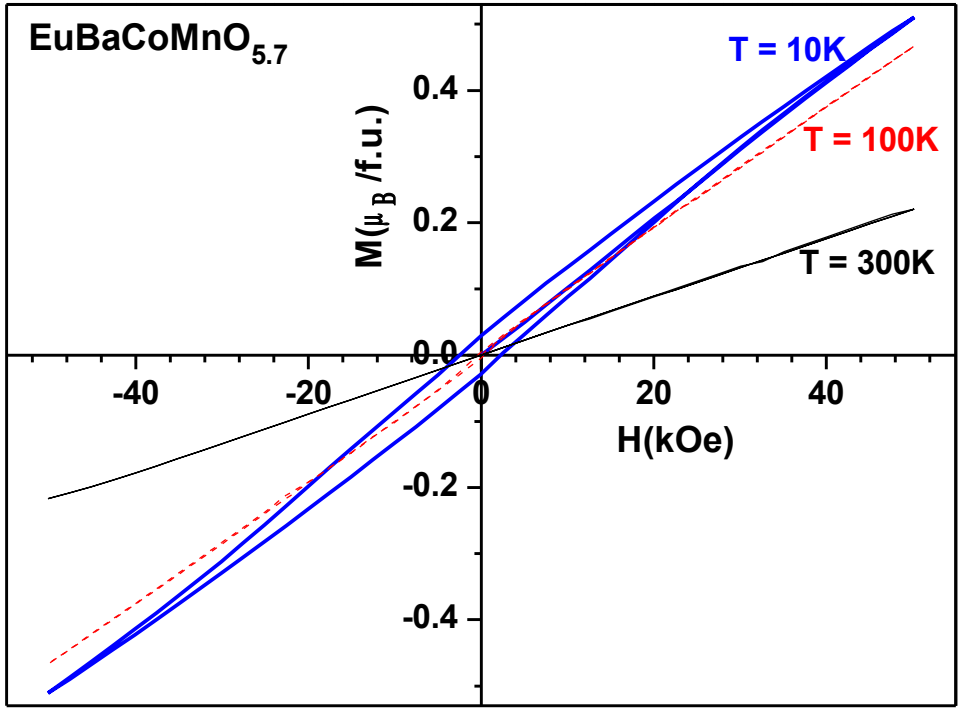

**FIG. 7.** Temperature dependent electrical resistivity, ρ, of LnBaCo$_{2-x}$(Mn)$_x$O$_{5+\delta}$ with Ln = Nd and Eu in the presence (solid symbol) and absence (open symbol) of magnetic field 7 Tesla. The inset figure shows magnetoresistance, MR(%), with the variation of temperature.

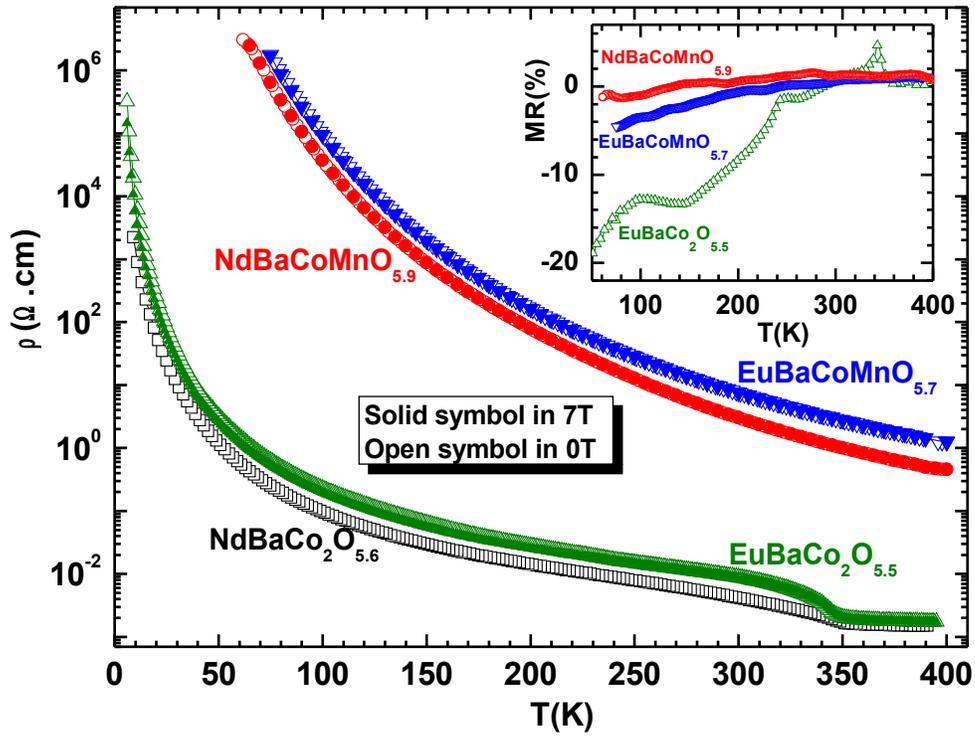